# A Proposal for an Open Logistics Interconnection Reference Model for a Physical Internet

Jean-Yves COLIN
LITIS, Université Du Havre
25 rue Ph. Lebon, BP 540
76058 Le Havre, France.
jean-yves.colin@univ-lehavre.fr

Hervé MATHIEU
ISEL, Université Du Havre
Quai Frissard
76063 Le Havre, France.
herve.mathieu@univ-lehavre.fr

Moustafa NAKECHBANDI
LITIS, Université Du Havre
25 rue Ph. Lebon, BP 540
76058 Le Havre, France.
moustafa.nakechbandi@univ-lehavre.fr

*Abstract*—This paper presents a New Open Logistics Interconnection (NOLI) reference model for a Physical Internet, inspired by the Open Systems Interconnection (OSI) reference model for data networks. This NOLI model is compared to the OSI model, and to the Transmission Control Protocol/Internet Protocol (TCP/IP) model of Internet. It is also compared to the OLI model for a Physical Internet proposed by Montreuil. The main differences between the presented NOLI model and all the other models named above are in the appearance of definitions of physical objects in different layers and not just the lowest one. Also, the NOLI model we present locates the containerization and de-containerization operations in the topmost layer, and not in the layer below as does the OLI model. Finally, the NOLI model is closer to the TCP/IP and OSI models than the OLI model, keeping the integrity of the Link Layer that the OLI model divides in two layers, and keeping separate the Session and Transport OSI Layers that the OLI model unites in just one layer.

## I. INTRODUCTION

The Internet network, a global system of interconnected computer networks, today links billions of devices worldwide, and is still growing. A network of networks, the Internet connects private, public, and academic heterogeneous networks of tiny to global sizes, using different technologies such as electronic (wireless or not) and optical means. It covers currently 46% of the earth population [7]. Information Systems enable the supply chain integration, specifically these last number of years. They made possible new production and distribution systems [1]. This trend is aiming toward a more integrated global supply chain. Though this trend is actual, it is not new and it was already pointed out in the 1960s, as a key area for future productivity improvements [2].

This integration trend has its limits and it becomes necessary to think over the logistic processes to be able to make them evolve. Thus, despite all the existing integration technologies being deployed in the supply chain, it is argued, for example in [11] that the transportation of physical goods is not as efficient, robust and sustainable as it could be. Concepts, tools and solutions developed in data networks provide interesting ideas to improve the efficiency and the sustainability of logistics networks.

However, today there is no practical solutions in logistics networks using the concepts developed in data networks, and it seems interesting to propose the idea of a Physical Internet. The idea of a worldwide Physical Internet dedicated to the moving, handling and storing of goods in standard containers, in a somewhat similar way the Internet is dedicated to the worldwide transportation and exchange of data, was first proposed in [8] and developed in several papers [9,6,12,13] and conferences (1st and 2d Conferences on Physical Internet, IPIC 2014 and IPIC 2015). As an aside, one can note that the idea of a Physical Internet is very different from the Internet-of-Things concept, which is about the idea that many physical objects, including but not limited to, cars, watches, refrigerators, pens, glasses, clothes, etc. can and should be connected to the Internet.

In this paper, we first present the formal OSI reference model for data networks, and then the TCP/IP reference model for the Internet. Next we present the OLI model of [9] for logistics networks, then we propose our own NOLI reference model for these networks. Finally, we compare our model to these models, and conclude with some final remarks.

## II. A PHYSICAL INTERNET

The idea of a Physical Internet [8] is to build a new, efficient worldwide logistics network, using all the researches results and ideas incorporated in the "electronic" Internet. This logistics network would connect different private and public heterogeneous logistics networks, to provide a more economically, environmentally and socially efficient, and sustainable mean to handle, move, store and use physical objects throughout the world. The Internet itself is based on the TCP/IP (Transmission Control Protocol/Internet Protocol) model. The goal of the TCP/IP model is to provide a set of protocols that separates the needs of communication applications from the specificities of the real networks used to transmit the data.

The separation thus realized makes the integration and the use of widely different real networks possible in an integrated global network. It also makes the design and development of applications easier. The OSI reference model, a more extensive, complex and refined formal model not specifically targeted for the Internet, was proposed independently [4]. We first present this model.

*A. The OSI reference model*

The Open Systems Interconnection reference model (OSI model) was initially proposed as a foundation and framework for the design of new data networks, protocols, and applications, and it was not intended for a specific network. Currently, it is used mainly as a pedagogical tool, but its influence is very important on the last versions of the TCP/IP model for Internet, and on new innovative network technologies. It is also almost always referred to, explicitly or implicitly, when describing data networks architectures, applications and protocols.

The OSI model is a conceptual model that characterizes and standardizes the communication functions of a telecommunication without concern for their underlying internal structure and technology. Its goal is the interoperability of diverse communication systems with standard protocols, which are the sets of rules that define how communication occurs in a network.

TABLE I. THE SEVEN LAYERS OF THE OSI REFERENCE MODEL

| Position in the OSI model | Layer Name | Role of the layer |
|---|---|---|
| 7 | Application Layer | Layer 7 is the point of contact of application with network services. |
| 6 | Presentation Layer | Layer 6 takes care of anything related to the presentation of data: format, encryption, encoding, compression, etc. |
| 5 | Session Layer | Layer 5 is in charge of the authentication, the initializing of a session, and its management and its closure. |
| 4 | Transport Layer | Layer 4 chooses the transmission protocol and prepares the data exchange from the starting location to the final destination. It splits data into multiple sequences (or segments). |
| 3 | Network Layer | Layer 3 establishes the logical connection between hosts. It deals with everything related to the address identification and routing in the network. It splits data into multiple packets. |
| 2 | Datalink Layer | Layer 2 establishes direct physical connections between hosts. It splits data into multiple frames. |
| 1 | Physical Layer | Layer 1 converts bit streams and physical transmissions of data on the media. It also defines the hardware and physical interfaces ( USB, DSL, Ethernet physical layer ) . |

The OSI model partitions a communication system into seven abstraction layers. Each layer serves the layer above it and is served by the layer below it. Table I presents a quick summary of the OSI layers and their functions.

The principle of a layer model is the following: if two nodes want to communicate (node A and node B), each N layer belonging to node A must communicate with the N layer belonging to node B. To do so, it gives the information to its N-1 layer, which will pass the information to its N-2 layer and so on, until the information reaches the physical network. Once the information arrives at the physical level, it is transmitted to the lowest layer of node B. The lowest layer of node B will then pass the information to the upper layers until it reaches the N layer of node B.

All models described in this article are based on this layer principle.

*B. The TCP/IP model of Internet*

TCP (at first called "Transmission Control Program") was initially designed around 1973 to support the development of the ARPAnet project of the US Department of Defense, and formally documented in RFC 675, "Specification of Internet Transmission Control Program", December 1974. The initial version had no layer. Due to growing design problems, TCP was renamed and divided into a TCP (for "Transmission Control Protocol") layer, and an IP (for "Internet Protocol") layer, around 1978, and formally documented in 1980 and 1981 with version 4 of TCP (RFCs 791, 792 and 793). The resulting model was named TCP/IP, and the suite of programs it included quickly became the core of ARPAnet, which later became the NSFnet, which finally evolved as the US core of what is now the Internet.

So the TCP/IP model as known today first had two layers, the Transport Layer, that is similar to Layer 4 (OSI model), and the Network Layer below it, that is similar to Layer 3 (OSI model). Thus its Transport Layer manages complete end-to-end transactions, while its Network Layer deals with the details of the routing of data through the networks [5].

TABLE II. COMPARISON BETWEEN THE OSI AND TCP/IP MODELS

| TCP/IP Layer Name | OSI Layer Name |
|---|---|
| Application | 7. Application |
| | 6. Presentation |
| | 5. Session |
| Transport | 4. Transport |
| Network | 3. Network |
| Network Access | 2. Data Link |
| Physical | 1. Physical |

Later, it became a four layers model, with the addition of an Application Layer, that included everything above the TCP Layer, and of a Network Access Layer (also called Network interface Layer, or Link Layer) below the IP Layer. The Network access Layer is very similar to the Data link Layer of the OSI model. The space left below the Network Access Layer (that occupies the same space as the first layer of the OSI model) is not formally documented, but is sometimes referred to as the physical layer. The main core of the TCP/IP model is its Transport Layer and its Network Layer (cf. Table II).

The TCP/IP model is still evolving today. The most visible change currently is the gradual replacement of the original IPv4 addressing system in its Network Layer by a new IPv6 addressing system.

*C. Comparison between physical networks and data networks.*

Although logistics networks include data too, there are some specificities to the physical networks when compared to data networks such as Internet.

There is a much bigger difference between the transported products inside the containers, and the containers themselves, than between data and data packets that are all sets of bits in data networks. Instead of just one kind of physical objects in data networks, there are actually three kinds of physical objects in physical networks: the physical means (as in data networks), the containers (that are just additional bits in data networks), and the goods (that are also just bits in data networks). This has many effects.

First, lost or damaged physical goods cannot be managed as easily as lost or damaged data packets. Data bits are just bits and can simply be sent again or dropped. Although there are additional delays involved when a data is sent again, the additional expenses are considered negligible.

However, physical goods may only be sent again directly if some full containers are available with the same exact (or sufficiently similar) product. If not, some new product must be re-ordered. In all cases, there are always additional expenses to be considered.

Even full containers that lost the name of their consignor (original sender) and of their consignee (final receiver) cannot simply be deleted as a data packet can be, and must be disposed of someway.

Additionally, some abilities of data networks, such as broadcasting or multicasting (a limited form of broadcasting) data, cannot easily be added to logistics networks, due to the physical nature of goods and of containers.

Next, physical means are much more complex resources than connected electronic communications, so they must be efficiently managed. Their number and availability are almost always limited one way or another (a ship is there or not, a crane may be limited to the lifting of light containers, etc.)

And the physical containers are limited resources too, especially the special ones: for example, reefers (refrigerated containers) are expensive and available in limited quantity only.

Also, the routing criteria in physical networks are much more complex than the usual distance or time criteria used when routing data packets. These criteria must include costs (local taxes and fees, strikes, etc.), and other factors, such as the security of the crew, of the carrier, and of the goods (affected by local political crisis, piracy, war, etc.). Also tradeoffs are possible: it is sometimes possible to accelerate things, for an additional price (overtime rates, increased wear on equipments, etc.)

However, computing times of routing algorithms in logistics networks tend to be much shorter than shipping times, compared to data networks. This is an advantage that may allow the use of much more complex algorithms when necessary.

## III. THE OLI MODEL

The OLI model is presented in [9] as a reference model for the concept of a Physical Internet that is proposed in [8]. Although it identifies the Internet as its guide, it is actually closer to the OSI model for data networks and to its seven layers architecture. Thus the OLI model includes seven layers too (cf. Fig. 1), but with some significant differences in the functionalities for some of its layers.

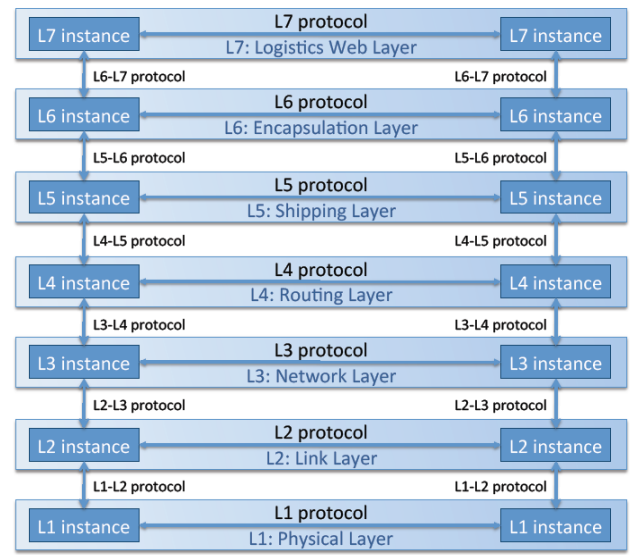

Fig. 1. The seven layers of the OLI model as presented in [9]

The seven layers of the OLI model (as proposed in [9]) are, presented below, from bottom to top.

*A. The Physical Layer*

The Physical Layer deals with all the operations concerning the elements of the Physical Internet. This includes the containers ($\pi$-containers), but also all the means ($\pi$-means) to move and to stock them (vehicles, stores, conveyors, etc.). The Physical Layer describes the physical interconnections of the Physical Internet: specifications of the $\pi$-containers, $\pi$-means monitoring, etc.

*B. The Link Layer*

The Link Layer monitors and tries to correct all unexpected events happening at the physical layer. This is being done by checking consistency between physical operations and their specifications.

*C. The Network Layer*

The Network Layer deals with the interconnectivity, integrity and interoperability of networks within the Physical Internet. It provides the means to route the $\pi$-containers across the network, and provides the quality of service requested by the Routing Layer. It also defines the composition of π-containers, and controls the flow of π-Containers within the network.

*D. The Routing Layer*

The routing layer is in charge of routing a set of π-containers from its source to its destination in an efficient and reliable manner. It defines the best path to use to route the π-containers according to networks status.

*E. The Shipping Layer*

The shipping layer provides all the means enabling the efficient and reliable shipping of sets of π-containers from consignors to consignees. It manages all the administrative aspects to reach this goal, and acknowledges (or not) the reception of the π-containers.

*F. The Encapsulation Layer*

The encapsulation layer assigns the products to their π-containers. It encapsulates products of a user in identified π-containers before accessing the Shipping Layer.

*G. The Logistics Web Layer*

The Logistics Web layer is the interface between the Physical Internet and the users of logistics services. It provides the users all the applications to exploit the Physical Internet.

## IV. THE NOLI MODEL

A lot of work was invested in the OSI model by its authors, and although it is not widely used in the design of actual data networks, its design is sound and is widely referred to when innovations, such as wireless communications, are introduced in real data networks. It is also close to the TCP/IP model of the Internet. The OLI model proposed above departs on several points from the OSI model, even if it keeps its seven layers architecture.

The NOLI we now present also keeps the seven layers of the OSI model.

*A. The Physical Handling Layer*

The Physical Handling Layer describes the physical characteristics of the π-means available to physically move the π-containers, such as ships, trucks, cranes, belt conveyors, etc. The long-distance conveyances and the local handling gears are considered to be at the same level in this model.

The Physical Handling Layer :

- Manages the states and localizations of these π-means (availability of a crane, etc.).
- Receives shipments of π-containers and the identification of the π-mean allocated to each shipment, from the Link layer.
- Manages the state (waiting, carried, done etc.) and localization of π-containers.
- Manages the scheduling of the π-containers on these π-means (to ensure that the maximum weight limit of a band conveyor is not exceeded, for example) or the mapping of the π-containers (which ones should be above on a container ship, etc. )
- Gives the orders to the π-means.
- Signals π-means problems (breakdowns, delays) to the Link layer.

This layer does not define the π-containers and their contents.

*B. The Link Layer*

The Link Layer manages the individual steps of movements of π-containers on π-means. A "step" is one individual point-to-point movement. The Link Layer receives blocks from the Network Layer with the starting and the ending location of each block.

The Link Layer divides and/or combine received blocks into several "shipments" and allocates a π-mean to each shipment to handle it for this step.

Although this may not be a physical move in some cases, the Link Layer also manages the handling of a block by a company/operator to another company/operator.

*C. The Network Layer*

The Network Layer receives loads of π-containers from the Transport Layer, with an initial starting and a final ending location for each load. The Network Layer divides and/or combines the received loads into "blocks".

The Network Layer computes and manages the routing of each block from its initial starting location to its final ending location. The Network Layer manages and maintains the data structures necessary to compute the best paths for the blocks.

*D. The Transport Layer*

The Transport Layer receives orders made of π-containers from the Order Layer, with an initial starting and a final ending location for each order. The Transport Layer divides and/or combines the received orders into "loads".

The Transport Layer manages the end-to-end trip of each load from its initial starting location to its final ending location. It checks that the final ending location can handle a load shipped there. It signals to the Order Layer the initial departure, the current location and the final arrival of each π-container. The Transport Layer ensures that deadlines are respected.

*E. The Order Layer*

The Order Layer receives sets of π-containers from the Container Layer, with an initial starting and a final ending location for each set. The Order Layer establishes the "dispatch note" associated to each π-container of each set. It also records priorities and deadlines of π-containers. The Order Layer divides and/or combines the sets into "orders" (according to deadlines, characteristics of π-containers, clients wishes such as sub-orders, etc.). The Order Layer checks the possible problems (for example, does the final ending location accepts dangerous material? etc.)

The Order Layer manages transactions. They can be simple complete orders, or more complex ones, such as sub-orders that may trigger intermediate payments if completed, etc. It signals damages to, or loss of, π-containers to the above Container Layer, and also received π-containers with no known consignor nor consignee.

*F. The Container Layer*

The Container Layer defines the physical characteristics of the π-containers allowed on the Logistics Network.

The Container Layer receives π-containers from the Product Layer, with contracts information. The Container Layer checks the physical integrity of received π-containers, and of the goods inside. The Container Layer combines the received π-containers into "sets". It also covers specialized nodes for the management of π-containers (empty containers, damaged containers testing, specialized containers maintenance). Finally, it manages received π-containers with no known consignor nor consignee.

*G. The Product Layer*

The Product Layer defines the possible products or goods that can be transported inside a π-container by the Physical Internet, and their characteristics.

The Product Layer fills empty π-container with the products. It establishes the contract for each filled π-container, and gives the filled π-containers and their contracts to the Container Layer. It receives filled π-containers from the Container Layer, checks their contracts and then empties them.

The Product Layer defines and establishes the different contracts and all their constraints, both for the consignor and the consignee, and for the Physical Internet.

A summary of al NOLI layers is presented in Table III. Examples of each layer of NOLI reference model are presented in Fig. 2.

TABLE III.  SHORT DESCRIPTION OF THE NOLI MODEL

| Position in the NOLI model | Layer Name | Role of the Layer |
|---|---|---|
| 7 | Product Layer | Defines the possible products or goods that can be transported inside π-containers. It fills the π-containers with the products and establishes the related contracts. |
| 6 | Container Layer | Defines the physical characteristics of the π-containers allowed on the Logistics Network. It will check the physical integrity of the π-containers and combine them into "sets" according to their characteristics. |
| 5 | Order Layer | Receives sets of π-containers from the Container Layer. It will create the orders according to the specified constraints (deadlines, client whishes, starting and destination point, etc.), and assigns the π-containers to the orders. |
| 4 | Transport Layer | Receives orders made of π-containers from the Order Layer. The transport Layer creates "loads" from the received orders, and manages the end-to-end trip for each load. |
| 3 | Network Layer | Receives loads of π-containers from the Transport Layer and creates "blocks" from the loads. The Network Layer defines a path across the network for each block. |
| 2 | Link Layer | Manages the individual steps (point-to-point movement) of π-containers on π-means. |
| 1 | Physical Handling Layer | Physical characteristics description of the π-means used to move the containers. |

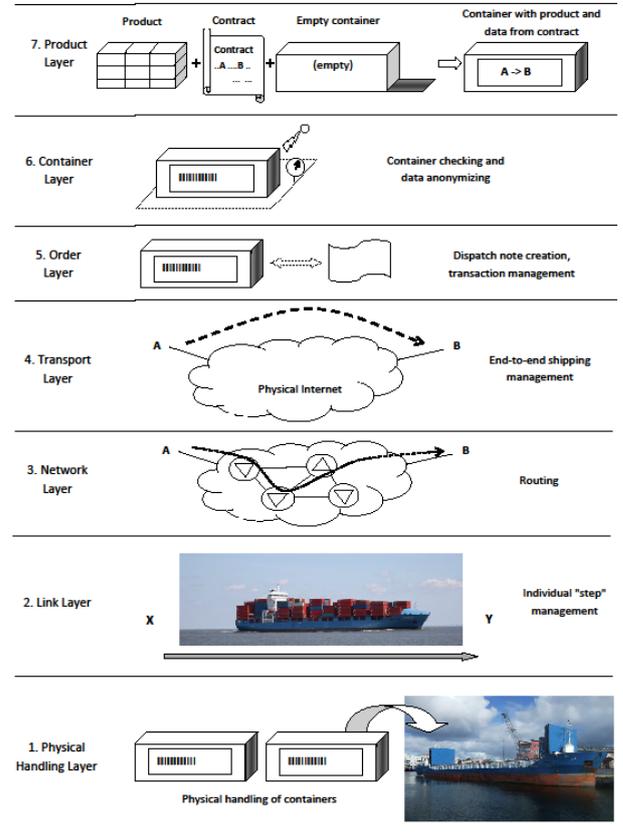

Fig. 2. Exemple of NOLI layers fonctionnalities

## V. DISCUSSION

Table IV presents the four models and gives a broad idea of the relationships between their respective layers. Please note that the boundaries are actually not as perfect as represented in this table: for example, the Transport Layers of the TCP/IP and of the OSI models are mostly identical, but the Transport Layer of the TCP/IP model includes a few functionalities related to transactions, that are defined and handled in the Session Layer of the OSI model.

The first difference between the other models and our NOLI model is in the location of the physical components definitions. The OLI model locates all definitions of the physical components in the lower layer, as do the OSI model and the TCP/IP model. One can note that the OSI and the TCP/IP models do locate them all in the lower layer, because the only true physical components of currecnt data networks are the carriers themselves indeed.

The carried objects (the data bits), and the enveloping objects (the data frames, or packets, etc.) that appear in, and are used by, the intermediate layers, are all just non-physical standard bits of data.

In logistics networks such as a Physical Internet however, the equivalent carried objects (the goods or cargo) and the enveloping objects (the containers) are true physical objects.

TABLE IV. COMPARISON BETWEEN THE LAYERS OF THE TCP/IP, OSI, OLI AND NOLI MODEL

| TCP/IP Layer Name (Internet) | OSI reference Model Layer Name | OLI Layer Name (Montreuil et al.) | NOLI Layer Name (Colin et al.) |
|---|---|---|---|
| Application | 7. Application | 7. Logistics Web | 7. Product |
| | 6. Presentation | 6. Encapsulation | 6. Container |
| | 5. Session | 5. Shipping | 5. Order |
| Transport | 4. Transport | | 4. Transport |
| Network | 3. Network | 4. Routing | 3. Network |
| | | 3. Network | |
| Network Access | 2. Data Link | 2. Link | 2. Link |
| Physical | 1. Physical | 1. Physical | 1. Physical Handling |

For this reason, we argue that there can be no single Physical Layer that would include all definitions of physical objects, and that the definitions of these physical objects in any Physical Internet model must be given in distinct layers, when they first appear.

Thus, the top-most Product Layer of our NOLI model defines the possible cargoes and their specificities. This includes the exact identification of the type of cargo, and its characteristics such as the fact that it is perishable or that it is fragile.

And in our NOLI model, the Container Layer below the Product Layer defines the characteristics of the π-containers. This includes specificities such as its size, or the fact that the π-container is a refrigerated one, for example.

We keep the π-means definitions in the lowest layer of NOLI model as does the OLI model, because the π-means are the physical logistics equivalents of the physical electronic components defined there in the OSI model, and it is logical to define this type of components in the lower layer.

Another difference between the OLI and our NOLI model is that the OLI model puts the containerization and de-containerization operations in the Encapsulation Layer below its topmost layer, while the NOLI model puts them in its topmost Product Layer.

We note that the topmost layer (Product Layer in NOLI) is conceptually the layer of the contents, and in logistics networks, the user is responsible for the direct handling of the contents. Everything below the top layer should have direct physical interaction with the π-containers only, and none directly with the contents, so the user should give and receive π-containers (and their attached contract statements) to the layer below him, ideally.

Next, the Transport Layer of the OLI model unites the OSI Transport Layer and the OSI Session Layer in just one layer.

We propose to keep the idea from the OSI model, that end-to-end transportation (the topic of these OSI layers) must be managed in two separate layers. We argue that one layer should be responsible for the administrative preparation and reception steps, so we call it the Order Layer, just below the Container Layer. And that the physical management of the end-to-end transportation needs one dedicated layer, below the Order Layer, more or less as it is the case in the OSI layer.

So we keep the OSI Transportation Layer name too for this lower layer. Also, we keep the OSI Network Layer, that is separated in the OLI model into a Routing Layer and a Network Layer, in one NOLI Network Layer that also includes the routing. We argue that the operations managed at this level, including the routing, belong in just one layer, as it is the case in the OSI model.

The decision to keep seven layers in any model may be considered somewhat arbitrary, and models with six (or less) layers or eight (or more) layers may be proposed in the future for logistics Networks. But, as we argued above in this paper, the OSI model seems to be a sound and internally consistent one. Any major difference with the OSI model must be justified, as is our proposal of having some physical components be defined outside the lower layer in our NOLI model.

VI. CONCLUSION AND FURTHER WORK

It is important to develop concepts and solutions to improve the logistics networks, in terms of efficiency and sustainability. Ideas developed in data networks for decades offer interesting solutions. It seems pertinent to adapt the layer models developed in data networks to define models that could be used in logistics networks.

In this paper, we presented a new Open Logistics Interconnection reference model for a Physical Internet, inspired by the Open Systems Interconnection reference model for data networks. We compared this NOLI model to the OLI model also for a Physical Internet, to the OSI model, and to the TCP/IP model of Internet.

Its main differences between the presented NOLI model and all the other ones are in the appearance of definitions of physical objects in several layers outside the lowest layer.

Also, the NOLI model we present locates the containerization and de-containerization operations in the topmost layer, and not in the layer below as does the OLI model.

The NOLI model is also closer to the TCP/IP and OSI models than the OLI model, keeping whole one layer that the OLI model divides in two layers, and not merging two other OSI layers that the OLI model unites in just one layer.

The next steps include the simulation of the NOLI model. The implementation of the NOLI layers implies the development of specific algorithms to perform each models function. We already worked in [3] on the balancing in logistics networks of mobile resources such as reefers.

ACKNOWLEDGMENTS

This work is supported by the Haute-Normandie Region : Projet CLASSE 2 "Corridors Logistiques : Applications à la vallée de la Seine et Son Environnement", France.